# Whistler Wave Plasma Production in CHS

T. Shoji [a], Y. Sakawa [a], C. Suzuki [b], K. Toi [b], R. Ikeda [b], M. Takeuchi [b], G. Matsunaga [c]
[a]*Nagoya University, Nagoya 464-8603, Japan,*
[b]*National Institute for Fusion Science, Toki 509-5292 Japan,*
[c]*Japan Atomic Energy Research Institute, Nka 311-0193, Japan*

Abstract

Rf plasma production in Whistler waves range of frequency $\omega/\omega_{ci} > 1$ is applied on the compact helical system (CHS) in a steady state operation with low toroidal magnetic field strength, where $\omega/2\pi$ and $\omega_{ci}/2\pi$ are rf and ion cyclotron frequencies, respectively. Maximum rf power and pulse width are is170kW and 10 msec at 9MHz, respectively. The line averaged plasma density $<n_e>$ as high as $2 \times 10^{18}$ m$^{-3}$ (He plasma) is produced in the region of toroidal magnetic field strength on magnetic axis B=100G~1.5kG and the electron temperature measured at plasma periphery is about 20eV. The wave magnetic field strength measured at the plasma periphery showed the evidence of the toroidal eigen modes of the Whistler waves. The additional long pulse microwave heating of 2.45GHz (>200msec) is applied and it sustains the plasmas while the microwave itself can not ignite plasmas when B< 600G.

Introduction

The rf plasma production was conducted on CHS by using Nagoya Type III antenna [1-2]. Efficient plasma production was obtained in the frequency range $\omega/\omega_{ci} < 0.8$ for B>5kG, where rf frequency $\omega/2\pi$=7.5-14MHz. Especially when $\omega/\omega_{ci} \sim 0.8$ on magnetic axis, efficient electron heating is observed. Ray tracing analysis of the Ion Bernstein wave excited at the antenna ($\omega/\omega_{ci} > 1$ at the antenna position) showed that the efficient electron heating occurs via Landau damping where the parallel wave number of the wave increases. For more higher magnetic field, slow wave contributes to produce plasmas. For the purpose of discharge cleaning, Alfven wave related studies and high beta related studies in helical systems, the plasma production in a low toroidal magnetic field in <kG range is preferable. Whistler wave discharge in MHz range of frequency have the notable feature of producing plasmas in such a low magnetic field where $\omega/\omega_{ci} > 1$ and the low phase velocity of the wave is expected to heat electrons to sustain plasmas.

Whistler wave plasma production

CHS[2-3] is a torsatron-type device having l=2/m=8 helical structure, and major and minor radius are 1m and 0.2m, respectively. Magnetic fields of CHS can be operated in steady state by SCR power supply when B<1.5kG. The rf frequency, maximum rf power and pulse width are 9MHz, 170kW and 10msec, respectively. Nagoya Type III antenna is used for the rf plasma production (Fig.1) and it is originally designed to flow the rf current along the magnetic field lines at plasma periphery for the excitation of Ion Bernstein waves. The antenna width is 15cm and the length along the magnetic field line is 65cm. 2mm microwave interferometer for averaged density $<n_e>$ and magnetic probe for wave field measurements are located 180$^0$ away from the rf antenna in toroidal direction. Movable Langmuir probe is installed in the same toroidal section where the antenna locates.

The Whistler wave propagation is examined by Ray tracing in the configuration where $\omega/2\pi$=13MHz, B=1.4kG ($\omega/\omega_{ci}$ =6 on magnetic axis), electron temperature $T_e$=100eV, electron density on axis is $1 \times 10^{18}$ m$^{-3}$ and initial refractive index along the toroidal magnetic field of the ray $N_{//}$ is 14. Figure 2 show that the ray propagates toroidally and absorbed by electrons via Landau damping. In the discharge process(electron heating) by the wave for the experimental condition, elastic and inelastic collisional damping also expected to contribute.

Figure 3(a) shows the dependence of rf produced helium (H$_e$) plasma density $<n_e>$ on B for the rf input power $P_{rf}$ =150kW and helium pressure $P_{He}$=5x10$^{-5}$ torr. In the low B region below 1kG, $<n_e>$ increases almost linearly with $P_{rf}$. $<n_e>$ of helium plasma as high as $2 \times 10^{18}$ m$^{-3}$ is produced and the electron temperature measured at plasma periphery is about 20eV by 150kW of the rf power. The plasma density produced here is factor ~5 smaller than the one at $\omega/\omega_{ci}$ ~0.8 (B>5kG, slow or Ion Bernstein waves) with the similar RF power. This may be attributable to the poor efficiency of the Whistler wave excitation for the Nagoya Type III antenna. Whistler wave dispersion relation is expressed as $c^2 k_{//}(k_{//}^2+k_{\perp}^2)^{1/2}=\omega\omega_{pe}^2/\omega_{ce}$, where $k_{//}$, $k_{\perp}$, $\omega_{pe}/2\pi$ and $\omega_{ce}/2\pi$ are parallel, perpendicular (to the magnetic field) wave number, electron plasma frequency and electron cyclotron frequency, respectively. It is conjectured that when B<500G, the wave number is mostly determined by the antenna, therefore $<n_e>$ is proportional to B. When B>500kG, $<n_e>$ tends to saturate, so the wave length become longer with B to have toroidal eigen modes. The dependence of $<n_e>$ on $P_{rf}$ for the helium plasma is shown in Figure 3(b). The plasma is ignited above 80kW and increases proportional to $P_{rf}$ under the conditions where B=800G and $P_{He}=5\times10^{-5}$ torr.

In order to study the wave propagation along the torus, wave magnetic field strength $B_{wave}$ is measured at the plasma periphery. Time evolution of $<n_e>$ and $B_{wave}$ are shown in Figure 4(a) and (b), respectively, here $P_{rf}$ =160kW and B=1kG. During the rf pulse, the plasma density changes in time and strong $B_{wave}$ peaks are observed at corresponding density $n_{peak}$. It is found that $n_{peak}$ increases with B as shown in Figure 5(a). Because of less damping of the wave expected from Landau damping and collisional damping with neutrals, the waves are expected to propagate around the torus as shown by the ray tracing calculation and form toroidal eigen modes. The toroidal eigen modes of the Whistler waves are calculated and shown in Figure 5(b), where the toroidal eigen mode number is N. The experimentally obtained $n_{peak}$ stays in the same range of the calculated toroidal eigen modes.

Effects of microwave heating(ECH) on rf plasmas
For the electron heating of Whistler wave plasma, the microwave power of 2.45GHz and 20kWx2 is used. The time evolution of $<n_e>$, the edge density $n_e(edge)$ ) and electron temperature Te(edge) at outmost magnetic surface measured by Langmuir probe are shown in Figure 6, here B= 613G and $P_{rf}$ =105kW. The plasma with peaked density profile on axis is produced initially by the rf power. At the last 4msec of the rf pulse, the microwave power of 18kW is applied on the plasma. Additional microwave power which is 17% of rf power causes the increase in average density of 20%, while $n_e(edge)$ increases three times as much as the rf plasma. After the rf power is shut off, the microwave power can sustain the plasma but $n_e(edge)$ jumps up and $<n_e>$ drops. This change in density implies that more flatter plasma profile than that of the rf plasma is produced by ECH. This is due to the fact that there is no density limit for the Whistler wave propagation, so the wave can penetrate at the center of the plasma but it is difficult for ECH(O or X modes) to penetrate deep inside the high density region. It is also shown that the plasma density can not start up for B< 600G when only the microwave power is applied. However by this rf start-up plasma can be sustained by ECH even when B<600G. Increase in $<n_e>$ with the microwave power under the fixed rf power is observed also above the cut-off density for 2.45GHz microwave.

Conclusion
Rf plasma production in Whistler waves range of frequency $\omega/\omega_{ci}$ >1 is applied on CHS. The line averaged He plasma density of $2\times10^{18}$ m$^{-3}$ is produced for the toroidal magnetic field strength on magnetic axis B=100G~1.5kG. The evidence of the toroidal eigen modes of the Whistler waves is observed. The additional long pulse microwave heating of 2.45GHz (>200msec) is applied and it can sustains the plasmas while the microwave itself can not ignite plasmas especially when B< 600G

Reference
[1] T. Shoji, K. Nishimura, et al., Nagoya Univ. Ann.Report 6(1989)1
[2] K. Nishimura, T. Shoji, , et al., Fusion Tech. 17(1990)86
[3] K. Nishimura, et al., IEA Stellarator Workshop on Future Large Devices (Oak Ridge, Tennessee, USA, 1990), Oak Ridge National Laboratory, Oak Ridge, Tennessee, Vol.2, (1990) .

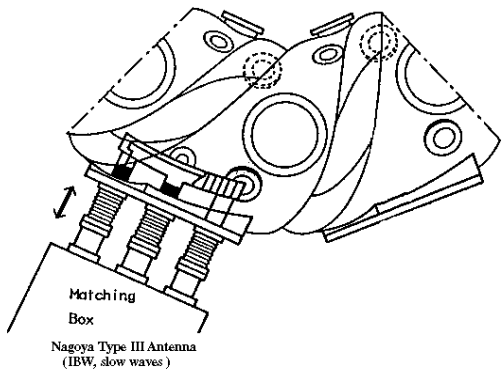

Fig.1 Nagoya Type III antenna

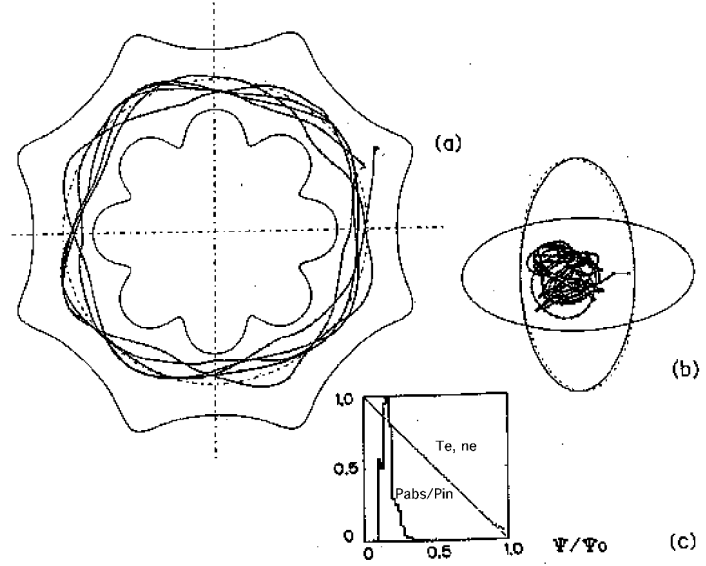

Fig.2 Ray trace of Whistler wave in CHS. (a)Toroidal cross section and (b) poloidal cross section. (c)Radial profile $\Psi/\Psi_0$ of $T_e$, $n_e$ and wave power absorption by electrons normalized by initial wave energy $P_{abs}/P_0$.

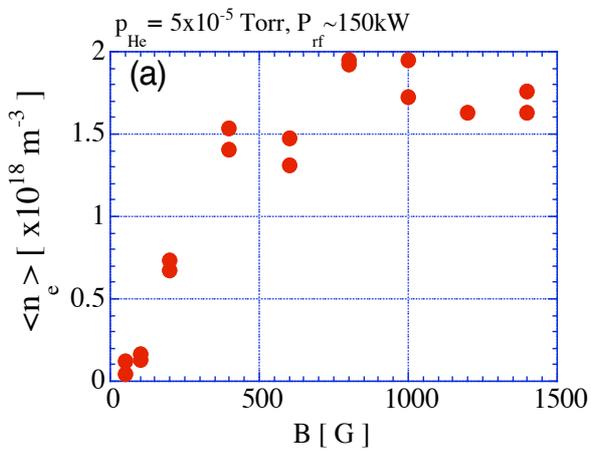
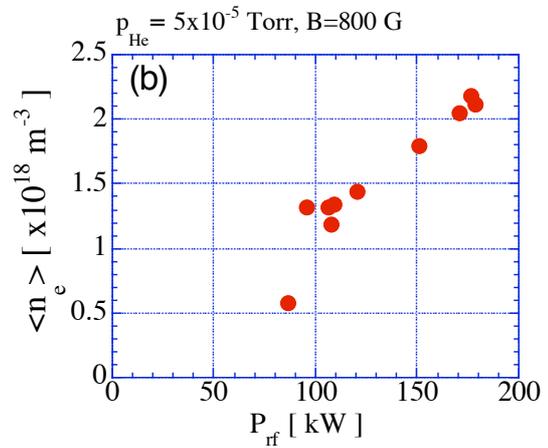

Fig.3 Toroidal magnetic field strength B (a) and rf power Prf (b) dependences on line averaged density $\langle n_e \rangle$.

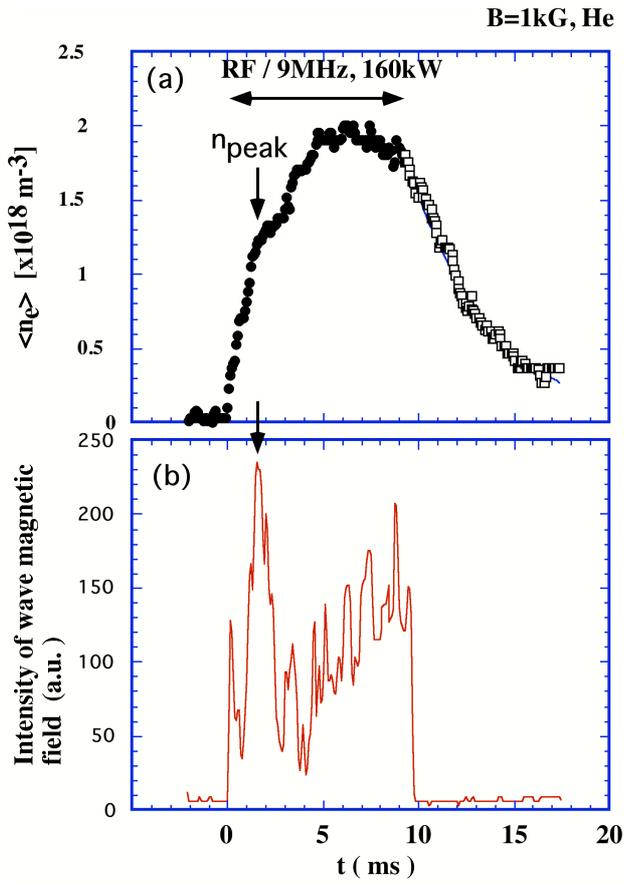
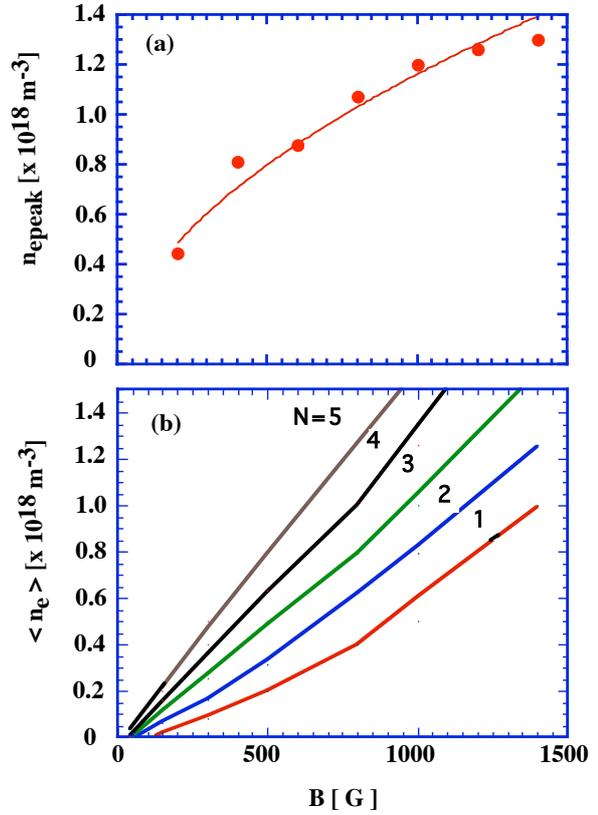

Fig.4 <$n_e$> (a) and wave magnetic field measured $180^0$ away from the antenna around the torus. Strong wave propagation is observed at $n_{peak}$.

Fig.5 (a) $n_{peac}$ obtained from the wave measurement as a function of B. (b) The relation between plasma density and B for Nth toroidal eigen modes of Whistler waves.

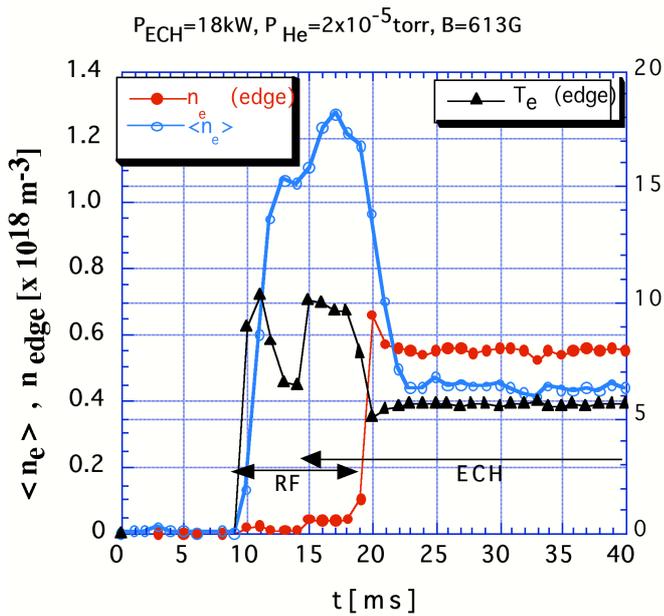

Fig.6 Time evolution of line averaged density, plasma edge density and electron temperature with and without ECH power for He plasma. f=9MHz.